\RequirePackage{rotating}
\RequirePackage{fix-cm}
\documentclass[smallextended]{svjour3}                     
\smartqed  
\usepackage{wrapfig}
\input{insbox}
\usepackage{graphicx}
\usepackage[misc,geometry]{ifsym}
\usepackage{ulem}
\usepackage{adjustbox}
\usepackage{framed}
\usepackage[utf8]{inputenc}
\usepackage{tikz}
\usepackage{pgfplots}

\definecolor{transfertoserver}{HTML}{D7191C}
\definecolor{database}{HTML}{FDAE61}
\definecolor{transfertoclient}{HTML}{ABDDA4}
\definecolor{rendering}{HTML}{2B83BA}
\usepackage{progressbar}
\usepackage{makecell}
\usepackage{booktabs}
\usepackage{tcolorbox}
\usepackage{amsmath}
\usepackage{amssymb}
\usepackage{paralist}
\usepackage{multirow}
\usepackage{xspace}
\usepackage{color}
\usepackage{xcolor}
\usepackage[graphicx]{realboxes}
\usepackage{ifthen}
\usepackage{url}
\usepackage{fancybox}
\usepackage{enumitem}
\usepackage{listings}
\usepackage{balance}
\usepackage{ifthen}
\usepackage{graphicx}
\usepackage{amsmath}
\usepackage[linesnumbered,boxruled]{algorithm2e}
\usepackage{algorithm2e}
\usepackage{tablefootnote}
\usepackage{comment}
\usepackage{float}
\usepackage{algorithmic}
\usepackage{dirtytalk}
\usepackage[]{natbib}
\usepackage{subfigure}
\usepackage{tcolorbox}
\definecolor{mygray}{gray}{0.6}

\usepackage[colorinlistoftodos]{todonotes}

\definecolor{chestnut}{rgb}{0.8, 0.36, 0.36}

\definecolor{chestnut}{rgb}{0.8, 0.36, 0.36}

\newlength\WIDTHOFBAR
\usepackage[colorlinks = true,
            linkcolor = black,
            urlcolor  = black,
            citecolor = black,
            anchorcolor = black]{hyperref}

\DeclareGraphicsExtensions{.pdf,.jpeg,.png}

	\newcommand{\RqOne}{RQ1: What is the reason of converting a discussion to an issue?}
	
	\newcommand{\RqTwo}{RQ2: What is the reason of converting an issue to a discussion?}
	
	\newcommand{\RqThree}{RQ3: How long does it take to raise a conversion intent?}
	
\begin{document}


\title{When Conversations Turn Into Work: A Taxonomy of Converted Discussions and Issues in GitHub}
\author{Dong Wang \Letter\and Masanari Kondo \and Yasutaka Kamei \and Raula Gaikovina Kula \and Naoyasu Ubayashi
}

\institute{
    \Letter~Corresponding author - Dong Wang\\
    Dong Wang, Masanari Kondo, Yasutaka Kamei, Naoyasu Ubayashi
    \at Kyushu University, Japan\\
    \email{\{d.wang, kondo,kamei, ubayashi\}@ait.kyushu-u.ac.jp}
    \and
    Raula Gaikovina Kula \at
    Nara Institute of Science and Technology, Japan\\
    \email{raula-k@is.naist.jp}\\
}

\date{Received: date / Accepted: date}
\maketitle
\begin{abstract}
Popular and large contemporary open-source projects now embrace a diverse set of documentation for communication channels.
Examples include contribution guidelines (i.e., commit message guidelines, coding rules, submission guidelines), code of conduct (i.e., rules and behavior expectations), governance policies, and Q\&A forum.
In 2020, GitHub released Discussion to distinguish between communication and collaboration.
However, it remains unclear how developers maintain these channels, how trivial it is, and whether deciding on conversion takes time.
We conducted an empirical study on 259 NPM and 148 PyPI repositories, devising two taxonomies of reasons for converting discussions into issues and vice-versa.
The most frequent conversion from a discussion to an issue is when developers request a contributor to clarify their idea into an issue (\textit{Reporting a Clarification Request --35.1\% and 34.7\%, respectively}), while agreeing that having non actionable topic  (\textit{QA, ideas, feature requests --55.0\% and 42.0\%, respectively}) is the most frequent reason of converting an issue into a discussion.
Furthermore, we show that not all reasons for conversion are trivial (e.g., not a bug), and raising a conversion intent potentially takes time (i.e., a median of 15.2 and 35.1 hours, respectively, taken from issues to discussions).  
Our work contributes to complementing the GitHub guidelines and helping developers effectively utilize the Issue and Discussion communication channels to maintain their collaboration.


\end{abstract}

\newpage

\section{Introduction}
\label{intro}

Contemporary open-source projects nowadays employ a plethora of communication channels to facilitate knowledge sharing and sustain the community around them~\citep{storey2016social}.
Complementary studies have shown that GitHub projects tend to adopt multiple communication channels and they are used to both capture new knowledge and update existing knowledge~\citep{tantisuwankul2019topological, vale2020relation}.
As part of a community, the management of communication and collaboration channels has led to the increase in the sophistication of documentation standards such as  contribution guidelines (e.g., commit message guidelines, coding rules, submission guidelines), code of conduct (e.g., rules and behavior expectations), governance policies, and Q\&A forum. 


Launched in 2020\footnote{\url{https://github.blog/2020-05-06-new-from-satellite-2020-github-codespaces-\\github-discussions-securing-code-in-private-repositories-and-more/}}, GitHub Discussion is a community forum that serves as an asynchronous communication channel.
The intended usage is for open-source communities, developer teams, and companies to ask questions, share ideas, and build connections with each other, all within the same GitHub platform. 
The first exploratory study by \cite{Hata2022GitHubDA} showed that Discussion is considered useful by developers and plays a crucial role in advancing the development of projects, uncovering several reasons for using GitHub Discussion mentioned in initial discussions (e.g., question-answering, community engagement, etc.).
Formally, Discussion is designed as a tool to share questions, ideas, conversations, requests for comment (RFC), resource planning, and community engagement.
This is different from the more traditional bug and code review systems such as GitHub Issue, which tracks executable pieces of work with a defined start and end point, including new features, fixing bugs, general updates, and tracking for epics and sprints, among other things.
Used together, GitHub claims that one benefit is that a developer can reference a Discussion in an issue as background and context for a piece of work, while converting an issue into a discussion could be due to the lack of information and decisions needed to complete a task.
Although GitHub provides  guidelines\footnote{https://resources.github.com/devops/process/planning/discussions/}, it remains unclear how developers maintain this intertwine between communication and actionable collaboration, how trivial is the conversion, and whether or not it takes time to decide on a conversion.

In this paper, we investigate the maintenance between communication and actionable collaboration by analyzing how developers decide between Discussion and Issue in real-world projects.
Hence, we conduct an empirical study to mine the conversions between them from 259 NPM repositories and 148 PyPI repositories.
Three research questions are formulated to guide this study:

\begin{itemize}
    \item \textbf{\RqOne}\\
    \emph{Motivation:} Although the prior work~\citep{Hata2022GitHubDA} explored the adoption of GitHub Discussion in terms of the usage and perception by developers, it is still unclear how contributors choose between the two communication channels.
    Specifically, Hata et al. studied the appearance of issue links and demonstrated that GitHub Discussions play a role in moving development forward by triggering new issues, but the intentions behind such triggers are not yet revealed.
    Answering this RQ would help the project to better maintain the communication channel and mentor newcomers to find a way to make contributions.
    \\
    \emph{Results:} Four reasons for suggesting a transition from discussions to issues are identified from a manual classification of a total of 331 samples, including Reporting a Bug, External Repository, Reporting a Clarification Request, and Reporting an Enhancement. Reporting a Clarification Request is the most common reason, with 35.1\% and 34.7\% of instances being classified for the NPM and the PyPI, respectively.
    \item \textbf{\RqTwo}\\
    \emph{Motivation:} As one strategy for moderating discussions, developers are allowed to convert an issue to a discussion. \cite{Hata2022GitHubDA} shows that around 18\% of discussions are converted from issues.
    However, the reason behind this conversion is not widely explored.
    Answering this RQ would help contributors further understand the proper position of the issue tracker system and avoid unnecessary burdens for engineering developers.
    \\
    \emph{Results:} Through a manual classification of a total of 433 samples, seven reasons are identified for converting issues to discussions.
    Non actionable topic is the most frequent reason, with 55.0\% and 42.0\% of instances being classified for the NPM and the PyPI, separately. Furthermore, Question-answering is the main non actionable topic.
    \item \textbf{\RqThree}\\
    \emph{Motivation:} We would like to explore whether deciding on a conversion from discussions to issues and vice versa takes time or not (i.e., discussion length and spent time).
    Answering this RQ would provide a potential future venue for the automatic classifier proposal to identify the appropriate topics between GitHub Issue and Discussion.
    \\
    \emph{Results:} The quantitative analysis shows that few posts are involved until the conversion intent is raised, i.e., mostly the median is one in both studied conversion kinds. However, raising the conversion intent could potentially take time. For instance, 15.2 hours and 35.1 hours (median) are taken until the conversion intent is raised from issues to discussions for the NPM and the PyPI, respectively.
\end{itemize}

Based on our empirical study results, we provide the following implications for the stakeholders.
For \textit{maintainers and contributors}, we find that our devised taxonomy complements the GitHub guidelines and further helps them decide when to contribute to
each channel and reduce unnecessary conversions.
Meanwhile, Discussion could be considered as a means to attract, and onboard potential new contributors.
We recommend that project maintainers should clearly state the submission rules in their README or contribution guidelines to avoid the inconsistent use of communication channels.
We also suggest contributors, especially newcomers, to start from the Discussion for the uncertainty problem (e.g., reporting potential bugs).
For \textit{researchers}, we provide the future direction of automatic classifier needs, including the detection of duplication between Issue and Discussion, the identification of non actionable topics from Issue, and bug-related topics from Discussion.


The remainder of this paper is organized as follows: Section \ref{sec:motivate} illustrates an example to motivate this study. 
Section \ref{sec:studied_datasets} describes the repository selection and the discussion extraction.
Section \ref{sec:approach} presents the approach to the proposed questions. 
Section \ref{sec:result} shows the research results.
Section \ref{sec:discussion} discusses the insights from our findings. 
Section \ref{sec:threat} discloses the threats to validity and Section \ref{sec:related} presents the related work.
Finally, we conclude the paper in Section \ref{sec:conclusion}.

\begin{figure}[]
\centering
\includegraphics[width=.95\linewidth]{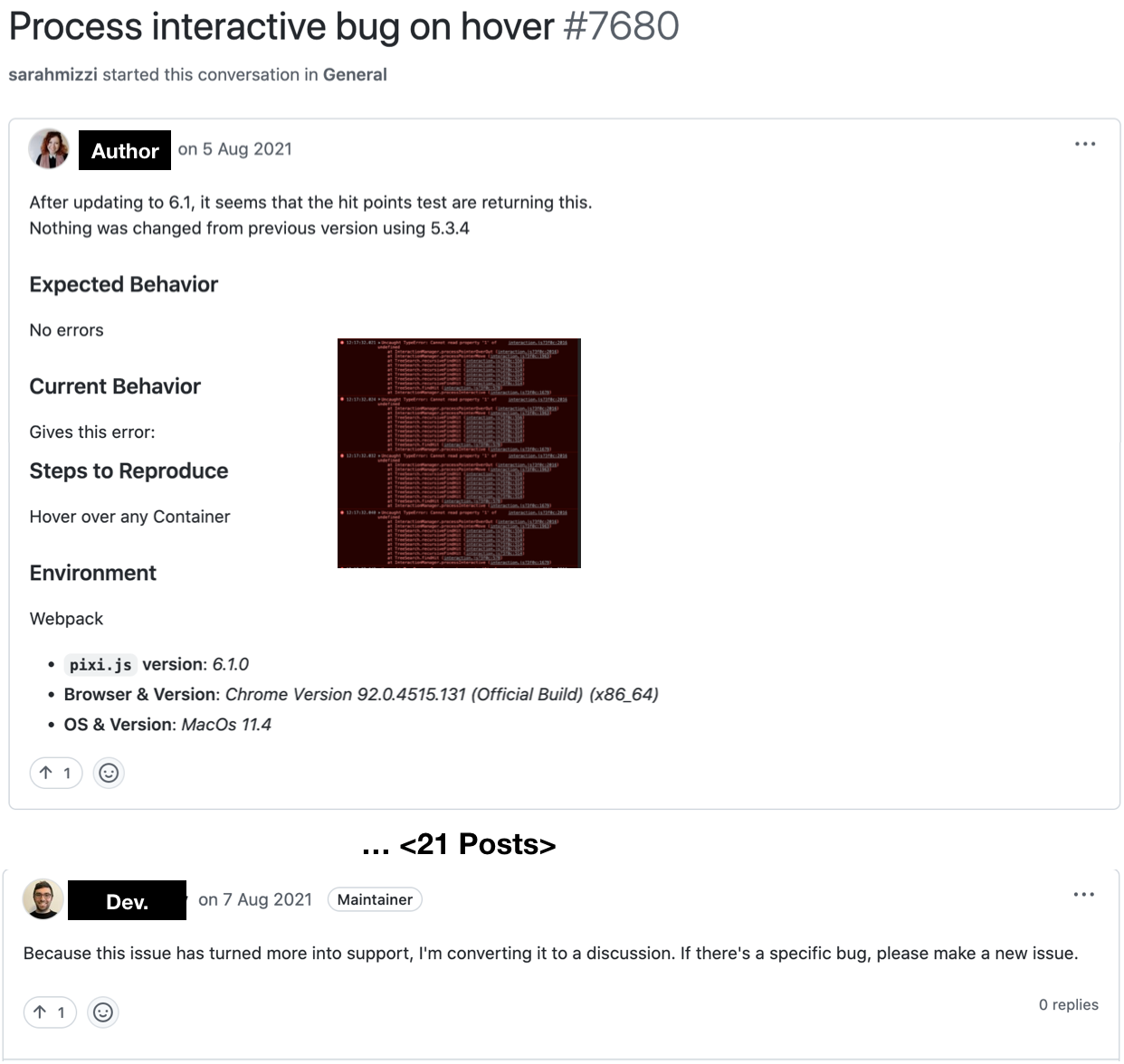}
\caption{GitHub issue is converted into Discussion thread (\texttt{pixijs \#7680}).}
\label{fig:example}
\end{figure}

\section{Motivating Example}
\label{sec:motivate}
As pointed out by the survey feedback in the work of ~\citet{Hata2022GitHubDA}, developers face the problem of topic duplication between Discussions and Issues.
Inspired by their work, we searched for some anecdotal evidence to understand the conversion process from Discussion to Issue and vice-versa.

Figure \ref{fig:example} shows an example from \texttt{pixijs \#7680} where a GitHub issue was converted into the Discussion channel. 
As shown in the figure, the issue was initially entitled as process interactive bug on hover.
To further confirm the bug or not, the author additionally provided the error message screenshot and the environment. 
Then, a conversation consisting of 22 posts was taken between four developers.
During the half-conversation, three developers tried to investigate the causes and claimed that they found a way to resolve the raised issues.
At the same time, they raised another potential problem around the pre-issue ``the issue seems to be bound to our codebase, not reproducible on a clean project'' and they further discussed this regarding the conflict of package versions.
At the end of this conversation, the maintainer suggested that this issue had turned more into support and raised an intention of converting it to a discussion thread.
In total, around two days were taken for this conversion process between the issue creation time (5 Aug 2021) and the conversion time (7 Aug 2021).
Although the work of~\citet{Hata2022GitHubDA} highlighted the challenge of choosing appropriate channels from the survey view, it is still unclear and lacks in-depth empirical analysis to reveal how the developers are using this new feature in real life. 
Inspired by this example, we hypothesize that (i) \textit{the conversion  process between Issue and Discussion may not be trivial}, and (ii) \textit{the conversion process may take a long period in terms of discussion length and time.}

\section{Studied Datasets}
\label{sec:studied_datasets}
In this section, we describe the process of preparing studied datasets, including repository selection and discussion extraction.

\begin{figure}[]
\centering
\includegraphics[width=.8\linewidth]{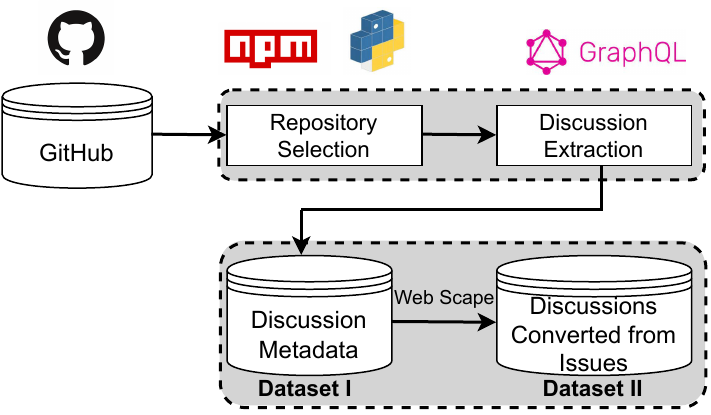}
\caption{The overview of dataset preparation.}
\label{fig:procedure}
\end{figure}

\paragraph{\textbf{Repository Selection.}} To answer our proposed RQs, we perform an empirical study on the NPM package  and PyPI ecosystems.
We selected the NPM and PyPI package ecosystems as (I) they are two of the largest package collections that are hosted on the GitHub platform and have been widely studied in the recent studies~\citep{10.1145/3106237.3106267,8894401,chinthanet2021lags}, (II) inspired by the work of~\citet{Hata2022GitHubDA}, 38\% of web libraries and frameworks (i.e., the highest proportion) have adopted the discussion beta feature.
Similar to the previous work~\citep{chinthanet2021lags}, we refer to the listing of NPM packages from the NPM registry and then matched them to the projects that are available in GitHub.
For the PyPI ecosystem, we rely on the open-source discovery service Libraries.io\footnote{{https://libraries.io/}} and the PyPI registry listing.
We assume that the more active and well-maintained package repositories are more likely to adopt the discussion feature.
Thus, we filtered and collected the repositories based on their contributor number (i.e., the number of contributors is larger than 100), resulting in 1,255 and 510 distinct repositories from the NPM and the PyPI, respectively.
Then, we use GraphQL API\footnote{https://graphql.org/} to determine whether or not these repositories have already adopted the Discussion feature. 
In the end, 263 NPM package repositories and 148 PyPI package repositories have introduced the GitHub Discussion, by the end of March 2022.

\paragraph{\textbf{Discussion Extraction.}} For the remaining 411 repositories, we then used GraphQL API to retrieve all discussions, including metadata such as discussion title, body, author, created time, and whether the discussion has selected answer or not.
For each post inside the discussion, we collected its body text, author, timestamp, and whether the post is an answer or not.
Moreover, for each post, we as well collected its nested replies, including reply body, reply time, and reply author.
After this, we were able to obtain 33,716 discussions from 259 NPM repositories (the rest of the four repositories have the Discussion feature but no discussions) with 68,695 individual posts and 69,593 replies in these posts, by April 2022. 
For the PyPI ecosystem, 12,662 discussions were retrieved from 148 repositories with 28,753 individual posts and their 29,001 replies.
We regard this dataset as \texttt{Dataset I}, as shown in Figure~\ref{fig:procedure}.
Since we would like to understand the characteristics and reasons of those discussion threads that are converted from the issues, we rely on the custom web scraper\footnote{https://github.com/sbaltes/github-retriever/} provided by~\citet{Hata2022GitHubDA} to determine whether the discussion thread was converted from an existing Issue.
Specifically, we provide the standard input as required (repository names and their owners) and the scraper (i.e., HTML parser) would automatically output the related attributes of discussion threads including the information of whether the thread was converted in the form of binary values.
Table \ref{tab:study_project} shows the statistic summary of our studied datasets.
As shown in the table, 5,689 discussion threads (16.8\%) and 3,337 discussion threads (26.4\%) are converted from the existing issues for the NPM and the PyPI, separately, denoted as \texttt{Dataset II}.

\begin{table}[]
\centering
\caption{Summary of Studied Datasets.}
\label{tab:study_project}
\resizebox{\textwidth}{!}{
\begin{tabular}{lrr}
\toprule
                                                 & NPM            & PyPI           \\ \midrule
Studied period                                   & \multicolumn{2}{c}{$\sim$ 2022.04}     \\
\# Package repositories                          & 259            & 148            \\
\# Discussion threads                            & 33,716         & 12,622         \\
\# Avg. Discussion threads per repositories      & 130            & 85             \\
\# Discussion posts                              & 68,695         & 28,753         \\
\# Discussion threads converted from Issues      & 5,689 (16.8\%) & 3,337 (26.4\%) \\
\# Avg. Discussion threads converted from Issues & 22             & 23             \\ \bottomrule
\end{tabular}}
\end{table}
 
\section{Approach}
\label{sec:approach}
In this section, we will describe the approach of our proposed RQs.

\subsection{RQ1 Analysis}
\label{sec:Rq2_approach}
To answer \emph{\RqOne}, we analyze a discussion that is suggested to be converted into an issue by looking at what is the conversion reason, using content analysis (one of the most broadly used qualitative data analysis methods)~\citep{stemler2000overview}.


\paragraph{\textbf{Identifying discussions that are converted to issues.}} Since there are no available datasets and tools or classifiers that were provided in the literature to be used, we first have to identify discussions that are suggested to be converted to issues.
To do so, 
we applied a semi-automatic method to identify such discussions from 46,388 discussion threads (33,716 and 12,622 discussion threads for the NPM and PyPI, respectively) and their posts (\texttt{Dataset I}), using a list of keywords.
To ensure the keyword is sufficient, we first manually inspected a group of 100 discussions that contain issue related links and picked up the potential indicator words.
Based on the observations and our knowledge, we refined the keyword and came up with the following keyword list: \texttt{\textless open, move, convert, create, please, transfer, submit, file, bug, issue\textgreater}, by taking case sensitive into account. 
Then we used this keyword list to match the discussion posts, resulting in 15,126 discussion threads (7,751 and 7,375 discussion threads for the NPM and PyPI, respectively) that at least contained one keyword in their posts.
Next, the first author manually validated these posts to identify the true positives where the discussion thread is suggested to be converted to an issue.
Finally, 595 discussion threads (374 and 221 discussion threads for the NPM and the PyPI, respectively) were collected as shown in Table~\ref{tab:representative}.

\paragraph{\textbf{Representative dataset construction.}} 
Similar to the prior work~\citep{WANG_emse, chouchen2021anti}, we then drew a statistically representative sample and the required sample size was calculated so that
our conclusions about the reasons of converted discussion threads would
be generalized to all discussion threads in the same bucket with a confidence level of 95\% and a confidence interval of 5.~\footnote{https://www.surveysystem.com/sscalc.htm} 
Thus, we randomly selected 190 and 141 discussion threads from the NPM and the PyPI that are suggested to be converted into issues to conduct the subsequent analysis, as shown in Table \ref{tab:representative}.

\begin{table}[]
\centering
\caption{Summary of Representative Dataset. \# of validated Dis. refers to the number of the discussions that are retrieved from the keyword list. \# of satisfied Dis. denotes the number of discussions that satisfy the criteria after the manual validation.
}
\label{tab:representative}
\resizebox{.9\textwidth}{!}{
\begin{tabular}{lrrrr}
\toprule
                          & \multicolumn{2}{l}{Dis. from issues (RQ1)} & \multicolumn{2}{l}{Dis. into issues (RQ2)} \\ \cmidrule{2-5} 
                          & NPM                  & PyPI                & NPM                  & PyPI                \\ \midrule
\# validated Discussions  & 7,751                & 7,375               & 1,156                & 523                 \\
\# satisfied Discussions  & 374                  &   221                  & 562                  & 438                 \\
\# representative samples & 190                  &   141                  & 228                  & 205                 \\ \midrule
Total                     & \multicolumn{2}{r}{331}                       & \multicolumn{2}{r}{433}                    \\ \bottomrule
\end{tabular}
}
\end{table}

\paragraph{\textbf{Manually Coding Reasons.}} We performed a manual analysis to investigate the reasons behind those discussions that are suggested to be filed as issues.
The manual analysis was conducted in multiple rounds, similar to the prior work~\citep{Hata:ICSE19}.
In the first round, the first two authors opened a round-table discussion on classifying twenty randomly selected discussion threads from the sample, and constructed an initial coding guide.
Note that to classify the reasons, we not only relied on the specific discussion comments, but also referred to the context of the whole discussion threads and sometimes tracked back to the opened issues, in order to obtain a comprehensive understanding.
To validate the coding guide and ensure that if there exist any missing codes, in the second round, the first two authors independently coded another twenty discussion threads. 
After this round, we found that the constructed coding guide can fit the sample and no new codes occurred.
Similar to prior work~\citep{WANG_emse, chouchen2021anti}, we then calculated the Kappa agreement of this iteration between two authors across four codes for these twenty discussions.
The score of the Free-marginal Kappa agreement is 0.80, implied as nearly perfect~\citep{mchugh2012interrater}.
Based on the encouraging results, the first author then coded the rest of the samples.


\subsection{RQ2 Analysis}
\label{sec:Rq1_approach}
To answer \emph{\RqTwo}, we analyze a discussion that is converted from an issue in terms of the reasons behind such conversion, through content analysis same to RQ1.

\paragraph{\textbf{Representative dataset construction.}} To understand what is the reason of this kind of discussion conversion, similar to RQ1, we perform a manual analysis on a statistically representative sample of our discussion dataset (\texttt{Dataset II}) where the discussion threads are automatically identified as converted from issues or not by the crawler tool.
Through an exploration of ten randomly selected samples, we observed that around 70\% of these discussions do not imply explicit reasons from their post context.
However, our study interest is to find the reasons and we would like to avoid the potential subjective threat.
Thus, we then applied a filter to retrieve those discussion threads that could probably provide the reasons.
To do so, we used the keyword \texttt{discussion} to narrow the data scope since we assume that the keyword \texttt{discussion} could indicate the discussion feature, resulting in 1,156 and 523 discussion threads for the NPM and the PyPI, separately.
Then, the first author manually inspected these discussion threads to validate whether or not their comments imply the conversion between discussions and issues.
We noticed that existing old issues can also be converted into discussion threads.
We argue that these existing old issues are out of our scope, as we are interested in the instances where issues are converted due to the adoption of the Discussion feature.
Therefore, we further excluded the converted issues that were submitted before the GitHub Discussion feature was firstly introduced (i.e., May 2020).
In the end, 562 and 438 discussion threads from the NPM and the PyPI satisfied the criteria.

To be consistent with the manual analysis in RQ1, we drew a statistically representative sample.
To reach the confidence level of 95\% and a confidence interval of 5, we then randomly sampled 433 discussion threads (228 and 205 threads for the NPM and the PyPI, respectively) from the bucket.
Table \ref{tab:representative} shows the summary of the studied dataset in RQ2.

\paragraph{\textbf{Manually Coding Reasons.}}We classify the reasons of converting issues to discussions, using statistically representative samples (433 discussion threads).
Our initial codes were informed by taxonomy to understand the reasons for using GitHub Discussions mentioned in initial discussions, such as question-answering, idea sharing, information resource building, and so on~\citep{Hata2022GitHubDA}.
We refer to their taxonomy since it is closely relevant to our study scope, and the taxonomy is validated by the systematic process.

To test whether or not the existing taxonomy can fit our cases, in the first iteration, we randomly selected twenty samples and classified them into the available codes between the first two authors.
After the classification, an open discussion was conducted between the two authors and we observed that the existing taxonomy cannot fit well.
We then refined the coding schema by modifying certain codes and adding new codes.
To validate our refined coding schema, another twenty samples were selected and the first two authors independently classified the samples.
After the second iteration, we found that there existed new codes that were not covered in our coding schema.
Then, an open discussion was held to discuss these new codes and further polish up our coding schema by emerging new codes.
To assure that there was no new code and further evaluated the coding schema, in the third iteration, we selected another twenty samples, and the first two authors independently coded them again.
After the third iteration, no new codes occurred and we found that the coding schema can fit well the samples.
In total, 60 samples were used to establish our coding schema of reasons of converting issues to discussions.
Note that our annotation is based on the whole issue conversation and the guidelines do not allow for multiple categories.
We then evaluated the inter-rater level of agreement between two raters across nine reasons, relying on the Kappa agreement.
The Kappa agreement score is 0.72 (i.e., substantial agreement).
Encouraged by this result, the first author then manually coded the rest of the 374 samples.
We classify those instances that do not fit the above codes into Others.
Then the above codes were merged into cohesive groups that can be represented by a similar subcategory (i.e., Question-answering, Idea sharing, and Feature request are merged into Non Actionable Topic), by conducting an open discussion among the four authors of this paper.

\subsection{RQ3 Analysis}
To answer \emph{\RqThree}, we perform a quantitative analysis on the manually labeled representative samples that are constructed from RQ1 and RQ2.


We define two metrics to conduct our statistical analysis, to understand the process of raising a conversion intent from discussion to issues and vice versa, as shown below:
\begin{itemize}
    \item \textit{Raising time (\# hours):} The duration that is from the time when the discussion or issue is submitted to the time when the post firstly suggests that the discussion or issue should be converted.
    \item \textit{Number of posts until the raising time:} The number of posts that are submitted until the first conversion intent is raised. Note that it includes the post in which the conversion suggestion is provided. 
    For instance, in the motivating example presented in Section 2, the twenty-second post suggests that the issue topic should be converted to a discussion.
    Thus, in this case, we count \textit{Number of posts until the raising time} as 22.
\end{itemize}
To assure that the post firstly raises the conversion intent, the first author manually validated 1,243 posts (740 and 503 posts for the NPM and the PyPI, respectively) and 1,938 posts (892 and 1046 posts for the NPM and the PyPI, respectively) from the studied discussions that are suggested to convert into issues (i.e., 331 samples in RQ1) or the discussions that are converted from issues (i.e., 433 samples in RQ2).
We then measure these two metrics for those labeled discussion threads that are either converted from issues or converted to issues.

Meanwhile, to further understand the effect of different reasons, we propose a null hypothesis that \textit{`Raising time and the number of posts until the conversion intent are significantly different among reasons of the conversion'}.
To statistically confirm the significant differences, we use the Kruskal-Wallis H test \citep{Kruskal:1952}. This is a non-parametric statistical test to be used when comparing two or more than two categories. In our study, there are four and seven main categories for RQ1 and RQ2, separately. 
The advantage of applying non-parametric statistical methods is that they make no assumptions about the distribution of the data~\citep{hecke2012power}.
In addition, we invoke a Mann-Whitney test~\citep{mann1947} to examine any significant difference in each pair category between NPM and PyPI ecosystems.

\section{Results}
\label{sec:result}
In this section, we present the results of the empirical study.
\subsection{Conversion from Discussions to Issues (RQ1)}
To answer RQ1, we analyze (I) the reasons of converting discussions to issues, and (II) the frequency of these reasons across our studied samples.
Below, we first provide representative examples for each reason type, and then we discuss the result of the frequency of reasons (Table~\ref{fig:rq2_reason} and Table~\ref{RQ2:frequency}).

\textbf{Taxonomy of Reasons.} Four reasons of converting discussions to issues are classified through our qualitative analysis, which is described in Section~\ref{sec:Rq2_approach}:

\underline{\textit{(I) Reporting a Bug.}} This category refers to the reason where the discussion post indicates that the discussion topic describes a bug.
In this category, the keyword ``bug'' is usually left on the post, suggesting that the submitted post is explicitly related to the bug.
For instance, in the Ex 1\footnote{https://github.com/prisma/prisma/discussions/10488}, one collaborator pointed out that the discussion topic (entitled as ``prisma generates creates a package-lock.json'') was indeed a bug and encouraged the author to open an issue in the repository, along with a directory structure sharing.\\

\fbox{\begin{minipage}{30em}
Ex 1\\
\textbf{Collaborator:} This doesn't sound right and it's definitely a \textbf{bug}. Could you please \textbf{open an issue} and share your directory structure in the monorepo and which directory you are running prisma generate from? Ideally, even a Git repository with minimal example that mimics your monorepo layout and triggers this bug. Thanks!
\end{minipage}}\\

\underline{\textit{(II) External Repository.}} This category emerges by grouping discussion threads in which the post indicates that the discussion topic should not be in the current repository, instead should be opened as an issue in another repository.
We observe that in this category, the appropriate repository that should be referred is always specifically provided by the collaborator. 
As shown in the Ex 2\footnote{https://github.com/facebook/docusaurus/discussions/6099}, a collaborator from the \texttt{docusaurus} repository posted a comment to make the author aware that the \texttt{openapi} plugin faces the incompatible issues with the latest chance, and suggested the author to open an issue in plugin related repository. 
\\

\fbox{\begin{minipage}{30em}
Ex 2\\
\textbf{Collaborator:} Hi, it seems the openapi plugin is using the docusaurus internals which is not compatible with the latest change. Please \textbf{open an issue} in \textbf{their repo} telling them that the constants have been moved to @docusaurus\/utils.
\end{minipage}}\\

\underline{\textit{(III) Reporting an Enhancement.}} This category denotes the reason where the post indicates that the idea or feature request or any warning that should be highlighted in the issue to be aware.
For example, in the Ex 3\footnote{https://github.com/eslint/eslint/discussions/14669}, an author submited a discussion to ask questions regarding \texttt{prefer-destructuring undestructurable}. 
One collaborator suggested the author to use slint-disabl, and also encouraged the author to file an issue to enhance this (i.e., ignore these cases by default).\\

\fbox{\begin{minipage}{30em}
Ex 3\\
\textbf{Collaborator:} you can just use eslint-disable. :)
it sounds a reasonable \textbf{enhancement} to ignore these cases by default (or behind an option). can you \textbf{file an issue}, thanks!
\end{minipage}}\\

\underline{\textit{(IV) Reporting a Clarification Request.}}
We define this category as the post indicates that the topic stated in the discussion thread is not clear or lacks of details, and an issue is suggested in order to better understand the problem (e.g., add reproduce examples) or better track the progress how the problem evolves.
We show the following two representative
examples to describe this reason.
In the Ex 4\footnote{https://github.com/gatsbyjs/gatsby/discussions/32147}, the collaborator was unsure about the proposed error message.
To further clarify this, the collaborator suggested the author to create an issue with a small reproduction.
In another example Ex 5\footnote{https://github.com/Automattic/mongoose/discussions/10516}, to investigate the problem reason, resulting from the mixed versions or not, the collaborator encouraged the author to open an issue.
\\

\fbox{\begin{minipage}{30em}
Ex 4\\
\textbf{Collaborator:} i'm \textbf{unsure} if we can get a better error message but could you maybe \textbf{create an issue} with a small reproduction?

\end{minipage}}\\

\fbox{\begin{minipage}{30em}
Ex 5\\
\textbf{Collaborator:} please \textbf{open an issue} and follow the issue template. it looks like you \textbf{might} have mixed versions of mongoose and mongodb.
\end{minipage}}\\


\begin{table}[]
\centering
\caption{Descriptions of reasons for converting discussions to issues.}
\label{fig:rq2_reason}
\begin{tabular}{lp{6cm}}
\hline
\textbf{Category}                & \textbf{Description}  \\ \toprule
(I) Reporting a Bug      & The comment indicates that the discussion topic describes a bug. \\ \midrule
(II) External Repository           & The comment indicates that the discussion topic should not be reported in the current repository, instead should be reported in another repository (e.g., dependent repository).\\ \midrule
(III) Reporting an Enhancement    & The comment indicates that an idea or a feature request or any warning is needed to be highlighted in the issue. \\\midrule
(IV) Reporting a Clarification Request               &  The comment indicates that to trigger more details and further confirm the problems (e.g., potential bugs), an issue is needed to follow up the discussion. \\ \bottomrule
\end{tabular}
\end{table}

\begin{table}[]
\centering
\caption{Frequency of reasons for converting discussions to issues.}
\label{RQ2:frequency}
\begin{tabular}{lrr}
\toprule
                           & \textbf{NPM}         & \textbf{PyPI} \\ \midrule
(I)  Reporting a Bug       & \textbf{44 (23.4\%)} & \textbf{32 (22.7\%)}    \\ \midrule
(II) External Repository  & \textbf{44 (23.4\%)} & \textbf{22 (15.6\%)}   \\ \midrule
(III) Reporting an Enhancement          & \textbf{34 (18.1\%)} & \textbf{38 (26.9\%)}  \\ \midrule
(IV) Reporting a Clarification Request & \textbf{66 (35.1\%)} & \textbf{49 (34.7\%)} \\ \bottomrule
\end{tabular}
\end{table}


\textbf{Frequency of Reasons.} We now examine what is the common reason of converting discussions to issues.
Table~\ref{RQ2:frequency} presents the distribution of the reason category in our studied NPM and PyPI repositories.
We observe that \textit{Reporting a Clarification Request} is the most common reason for both package ecosystems, with 66 (35.1\%) and 49 (34.7\%) discussion threads being classified, separately.
Such a result indicates that the discussions that should be converted to issues are more likely to result from uncertain discussion topics (e.g., potential bugs), which request further clarification from the issue tracker support.
The following two reasons are the second most popular for the NPM repositories, i.e., \textit{Reporting a Bug} and \textit{External Repository}.
44 discussion threads were manually classified for both cases, accounting for 23.4\% of instances.
While, for the PyPI repositories, the following frequent reason is \textit{Reporting an Enhancement}, with 26.9\% of instances being classified.

\begin{tcolorbox}[colback=gray!5,colframe=gray!75!black,title= RQ1 Summary]
Four reasons are classified behind converting discussions to issues, including Reporting a Bug, External Repository, Reporting a Clarification Request, and Reporting an Enhancement.
Reporting a Clarification Request is the most common reason for the two studied ecosystems (NPM and PyPI), accounting for 35.1\% and 34.7\% of instances, respectively.
\end{tcolorbox}

\subsection{Conversion from Issues to Discussions (RQ2)}
To answer RQ2, we analyze (I) the reasons of converting issues to discussions, and (II) the popularity of the classified reasons.
We now illustrate the reasons using representative instances and discuss the frequency results (Table~\ref{RQ1:result} and Table~\ref{RQ1:frequency}).

\textbf{Taxonomy of Reasons.} Nine reasons of converting issues to discussions are classified through our qualitative analysis, which is described in Section~\ref{sec:Rq1_approach}:

\underline{\textit{(I) Non Actionable Topic.}} This category relates to the reason where the topic proposed in the issue is not actionable and does not fit the issue scope. 
The category includes three sub-codes: \textit{Question-answering}, \textit{Feature requests}, and \textit{Idea sharing}.
For example, in the Ex 1\footnote{https://github.com/aws-amplify/amplify-js/discussions/8106}, the author submitted an issue to ask for \textit{``DataStore: How to handle a partial sync up to AppSync?''}.
While the maintainer left the comment and considered this topic was more of a general question that should be in the discussion area. Thus, we classify its reason into \textit{Question-answering}.
In another example Ex 2\footnote{https://github.com/grafana/grafana/discussions/46356}, the author proposed an issue to report a problem regarding query result sharing (i.e., \textit{Unable to select Query X in pane}), attached with the screenshots showing unexpected results.
However, a collaborator pointed out that this issue should be converted into the discussion as a feature request.

\fbox{\begin{minipage}{30em}
Ex 1\\
\textbf{Maintainer:} Hey Thanks for raising this! After reading over this issue, it appears to be \textbf{more of a general question} or topic for discussion and will be labeled as such. This is to differentiate it from the other types of issues and make sure it receives the attention it deserves.\\
...
\end{minipage}}\\

\fbox{\begin{minipage}{30em}
Ex 2\\
\textbf{Collaborator:} Hello @\textless username\textgreater, thanks for reporting this.
The feature is working as intended as it allows to share all the results in a panel, also as described in the docs.
I do agree however that it may be an interesting feature.
I'm converting this to a discussion where we \textbf{track feature requests}.
\end{minipage}}\\

\underline{\textit{(II) Invalid Issues.}} This category is merged by two cohesive codes (i.e., \textit{Lack of description information}, \textit{Not follow the template}), referring to the reason where the reported issue lacks of sufficient information for other developers to investigate. The example Ex 3\footnote{https://github.com/logaretm/vee-validate/discussions/3723} illustrates a scenario regarding \textit{Lack of description information}.
As we can see, the maintainer can not understand the issue fully with only provided code and suggested the author to create a minimal live example.
Furthermore, this issue was then moved to discussion due to its invalidity.

\fbox{\begin{minipage}{30em}
Ex 3\\
\textbf{Maintainer:} Please \textbf{create a minimal live example} on codesandbox, I can't guess the issue from this code alone. Also moved this to discussions since this doesn't satisfy an ``issue'' report.
\end{minipage}}\\

\underline{\textit{(III) Not a Bug.}} This denotes the reason where the author proposes a bug report, while the developer does not agree that it should be a bug. 
For instance, in the example Ex 4\footnote{https://github.com/keycloak/keycloak/discussions/8988}, the author followed the issue template including unexpected behavior and reproduction steps to report a potential bug related to password reset.
However, the maintainer did not agree with this and argued that this was not a bug, instead due to design and changed it to a discussion.

\fbox{\begin{minipage}{30em}
Ex 4\\
\textbf{Maintainer:} Changing this to a discussion as this is \textbf{not a bug} and is by design.\\
Recover password is used in scenarios when a user has forgotten the password, as such should not invalidate other sessions. In fact that is pretty common practice.
If a user suspect the account has been compromised they will update the credentials through the account console, which gives the option to logout existing sessions.
\end{minipage}}\\

\underline{\textit{(IV) Further Discussion.}} This category refers to the reason where the issue requires further confirmation or additional feedback from GitHub Discussion.
As shown in the Ex 5\footnote{https://github.com/serialport/node-serialport/discussions/2287}, to further validate the problem cause of \textit{``SerialPort.write(): callback never called''}, the maintainer moved the issue to the discussion system.

\begin{table}[]
\centering
\caption{Descriptions of reasons for converting issues to discussions.}
\label{RQ1:result}
\begin{tabular}{lp{6.5cm}}
\hline
\textbf{Category}     & \textbf{Description} \\ \toprule
(I) Non Actionable Topic   &  \multirow{4}{5cm}{The comment indicates that the topic proposed in the issue is not actionable and does not fit the issue scope.} \\
Question-answering      &     \\
Idea sharing            &    \\
Feature request         &     \\ \midrule
(II) Invalid Issues     & \multirow{3}{5.5cm}{The comment indicates that the reported issue lacks of sufficient information for the developers to investigate.} \\
Lack of description     &    \\
Not follow the template &   \\ \midrule
(III) Not a Bug               & The comment indicates that the bug proposed by the author does not get recognized by developers. \\ \midrule
(IV) Further Discussion      &  The comment indicates that the issue
requires further confirmation or additional feedback from GitHub discussion. \\ \midrule
(V) Already Fixed           &  The comment indicates that the issue has already been raised in
either the issue lists or the discussion thread. \\ \midrule
(VI) External Repository   & The comment indicates that the discussion topic should not be reported in the current repository, but instead should be reported in another repository (e.g., dependent repository)\\ \midrule
(VII) Information Storage     &  The comment indicates that the issue contents should be kept as a reference for the community.   \\ \bottomrule
\end{tabular}
\end{table}

\begin{table}[]
\centering
\caption{Frequency of reasons for converting issues to discussions.}
\label{RQ1:frequency}
\begin{tabular}{lrr}
\toprule
                          & \textbf{NPM}          & \textbf{PyPI}        \\ \midrule
(I) Non Actionable Topic  & \textbf{121 (55.0\%)} & \textbf{86 (42.0\%)} \\
Question-answering        & 76 (34.5\%)  & 51 (24.8\%) \\
Idea sharing              & 24 (10.9\%)  & 20 (9.8\%)  \\
Feature request           & 21 (9.5\%)   & 15 (7.3\%)  \\ \midrule
(II) Invalid Issues       & \textbf{19 (8.6\%)}   & \textbf{34 (16.6\%)} \\
Lack of description       & 17 (7.7\%)   & 32 (15.5\%) \\
Not follow the template   & 2 (0.9\%)    & 2 (1.0\%)   \\ \midrule
(III) Not a Bug           & \textbf{45 (20.5\%)}  & \textbf{38 (18.4\%)} \\ \midrule
(IV) Further Discussion   & \textbf{10 (4.5\%)}   & \textbf{17 (8.3\%)}  \\ \midrule
(V) Already Fixed         & \textbf{7 (3.1\%)}    & \textbf{6 (2.9\%)}   \\ \midrule
(VI) External Repository & \textbf{13 (5.9\%)}   & \textbf{7 (3.4\%)}   \\ \midrule
(VII) Information Storage & \textbf{5 (2.2)}      & \textbf{17 (8.3\%)}  \\ \bottomrule
\end{tabular}
\end{table}

\fbox{\begin{minipage}{30em}
Ex 5\\
\textbf{Maintainer:} I'm going to convert this to our new discussion system until we \textbf{confirm} some sort of bug
\end{minipage}}\\

\underline{\textit{(V) Already Fixed.}} This category denotes the reason where the comment indicates that the issue has already been raised in either the issue lists or the discussion thread. As shown in the Ex 6\footnote{https://github.com/gatsbyjs/gatsby/discussions/31283}, the maintainer suggested that the pitfalls related to build has been addressed by adding a doc page and then converted the existing issue into a discussion thread.

\fbox{\begin{minipage}{30em}
Ex 6\\
\textbf{Maintainer:} Since we \textbf{added} a doc page about these pitfalls and there's \textbf{nothing to ``fix''} here I'll move this to discussion :)
\end{minipage}}\\

\underline{\textit{(VI) External Repository.}} This reason refers to the case where the comments indicate that the issue is not raised in the appropriate place. For instance, in the Ex 7\footnote{https://github.com/facebook/create-react-app/discussions/11405}, the author proposed an issue concerning task tracker app in the create-react-app repository. However, the collaborator considered that the issue was from CRA tool and further this issue was converted.

\fbox{\begin{minipage}{30em}
Ex 7\\
\textbf{Collaborator:} Hi @$<$username$>$, I converted this issue into an discussion as the issues are more for \textbf{CRA tooling} internal issues.
\end{minipage}}\\

\underline{\textit{(VII) Information Storage.}} It means that the issue is converted since the comment indicates that the existing issue should be kept as a reference for the community. For example, in the Ex 8\footnote{https://github.com/date-fns/date-fns/discussions/2841}, the author raised a problem regarding react-native formatting and seeked for the solution. One collaborator provided a couple of solutions and later moved this issue into a discussion for other developers to find it easily.

\fbox{\begin{minipage}{30em}
Ex 8\\
\textbf{Collaborator:} Going to move this to a discussion so it's \textbf{easier for people to find}.
\end{minipage}}\\




\renewcommand\cellset{\renewcommand\arraystretch{0.3}%
\setlength\extrarowheight{0pt}}

\textbf{Frequency of Reasons.} Table \ref{RQ1:frequency} shows the reason frequency of converting issues to discussions.
As shown in the table, we observe that \textit{Non actionable topic} is the most common reason category within two studied package ecosystems, with 121 instances (55.0\%) and 86 instances (42.0\%) being classified for the NPM and the PyPI, separately.
Upon closer look, \textit{Question-answering} is the main non actionable topic, accounting for 34.5\% and 24.8\% of the occurrences of the \textit{Not actionable topic} category for the NPM and the PyPI, respectively.
The second most frequently occurring reason category is \textit{Not a bug} (20.5\% and 18.4\% for the NPM and PyPI, respectively), where the identified bug by the author is not recognized by other developers.
Least frequently occurring pattern includes the \textit{Already fixed} reason, with 3.1\% and 2.9\% of instances being classified for two ecosystems.

\begin{tcolorbox}[colback=gray!5,colframe=gray!75!black,title= RQ2 Summary]
Seven reasons are identified for converting issues to discussions.
Non actionable topic is the most frequent reason, with 55.0\% and 42.0\% of instances being classified for the NPM and the PyPI, separately.
Furthermore, Question-answering is the main non actionable topic.
\end{tcolorbox}

\subsection{Analysis of the Conversion Process (RQ3)}
\begin{figure}[]
    \centering
    \subfigure[Raising time of the conversion intent.]{\includegraphics[width=0.8\textwidth]{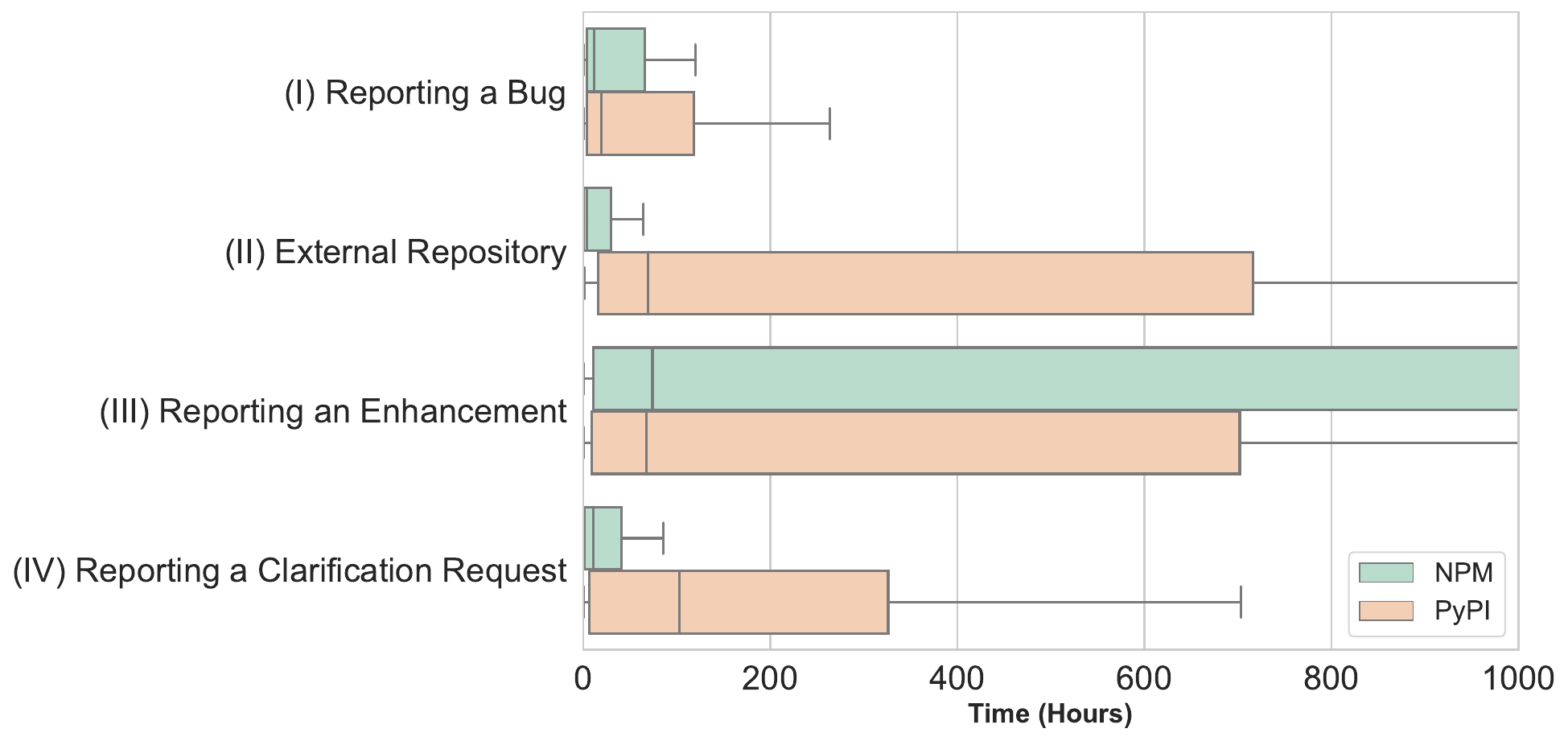}}
    \subfigure[Number of posts until the raising time.]{\includegraphics[width=0.8\textwidth]{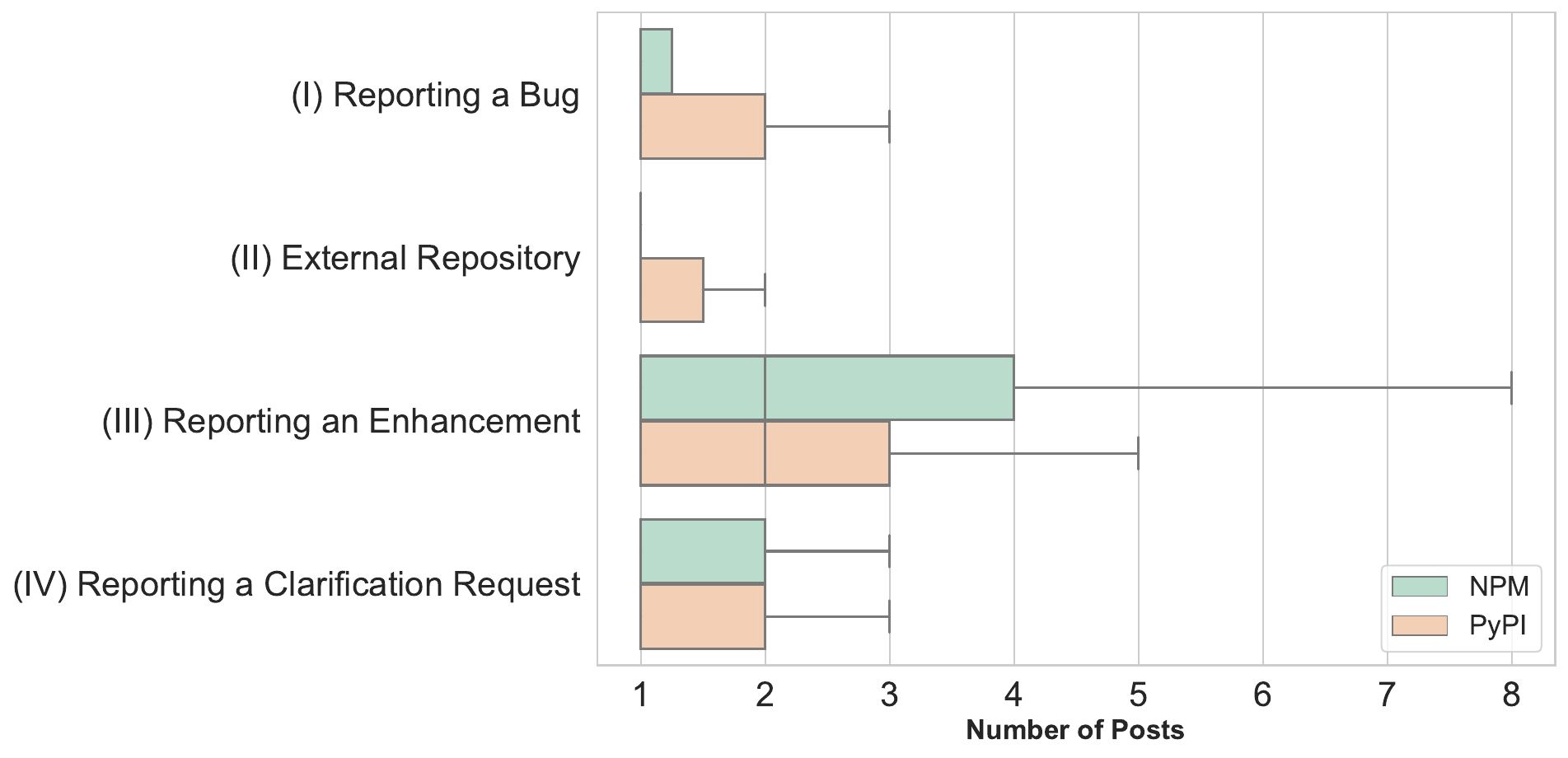}}
    \caption{Discussions that are converted into issues: (I) raising time and (II) number of posts until the raising time.}
    \label{fig:RQ3_a}
\end{figure}

\begin{figure}[]
    \centering
    \subfigure[Raising time of the conversion intent.]{\includegraphics[width=0.8\textwidth]{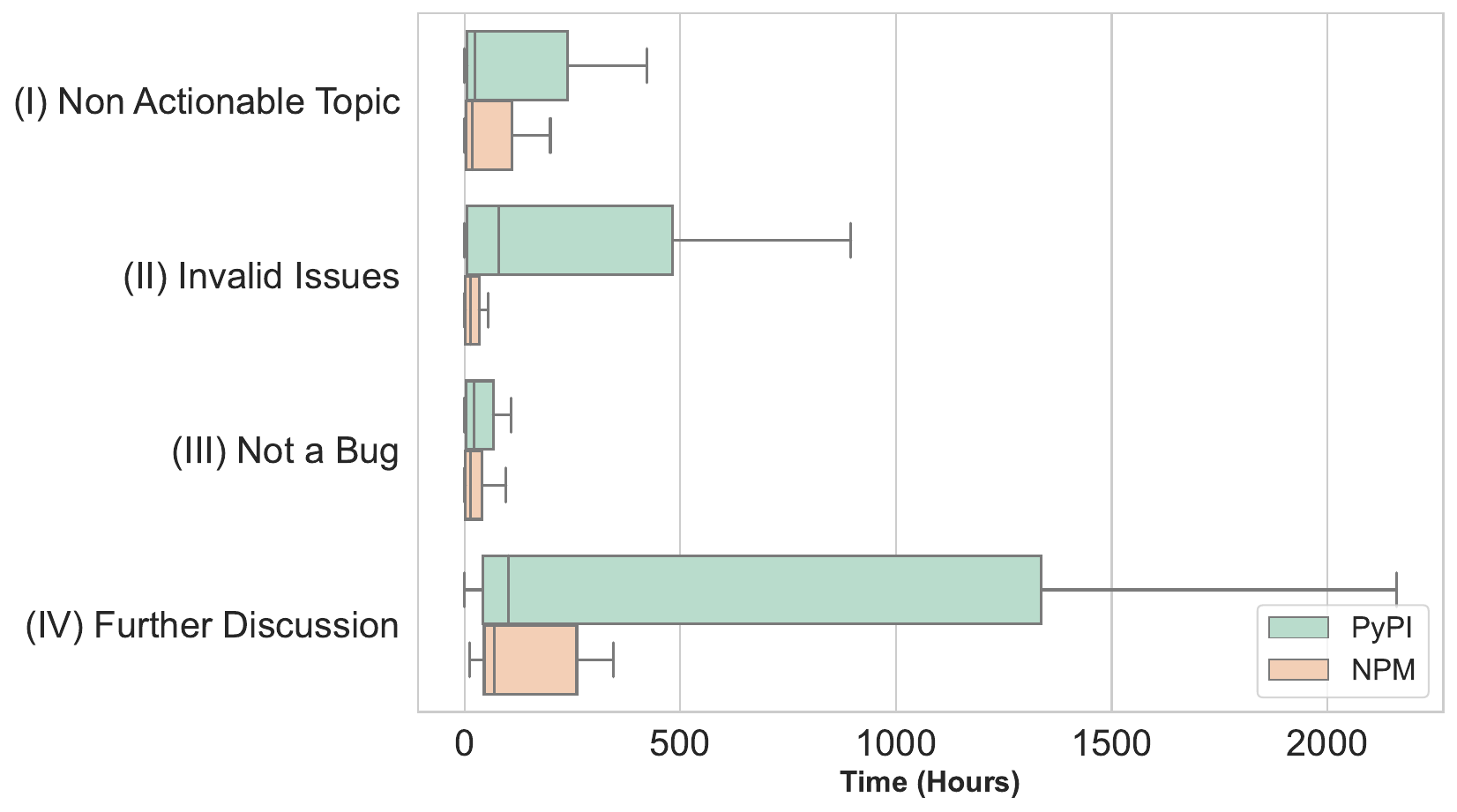}}
    \subfigure[Number of posts until the raising time.]{\includegraphics[width=0.8\textwidth]{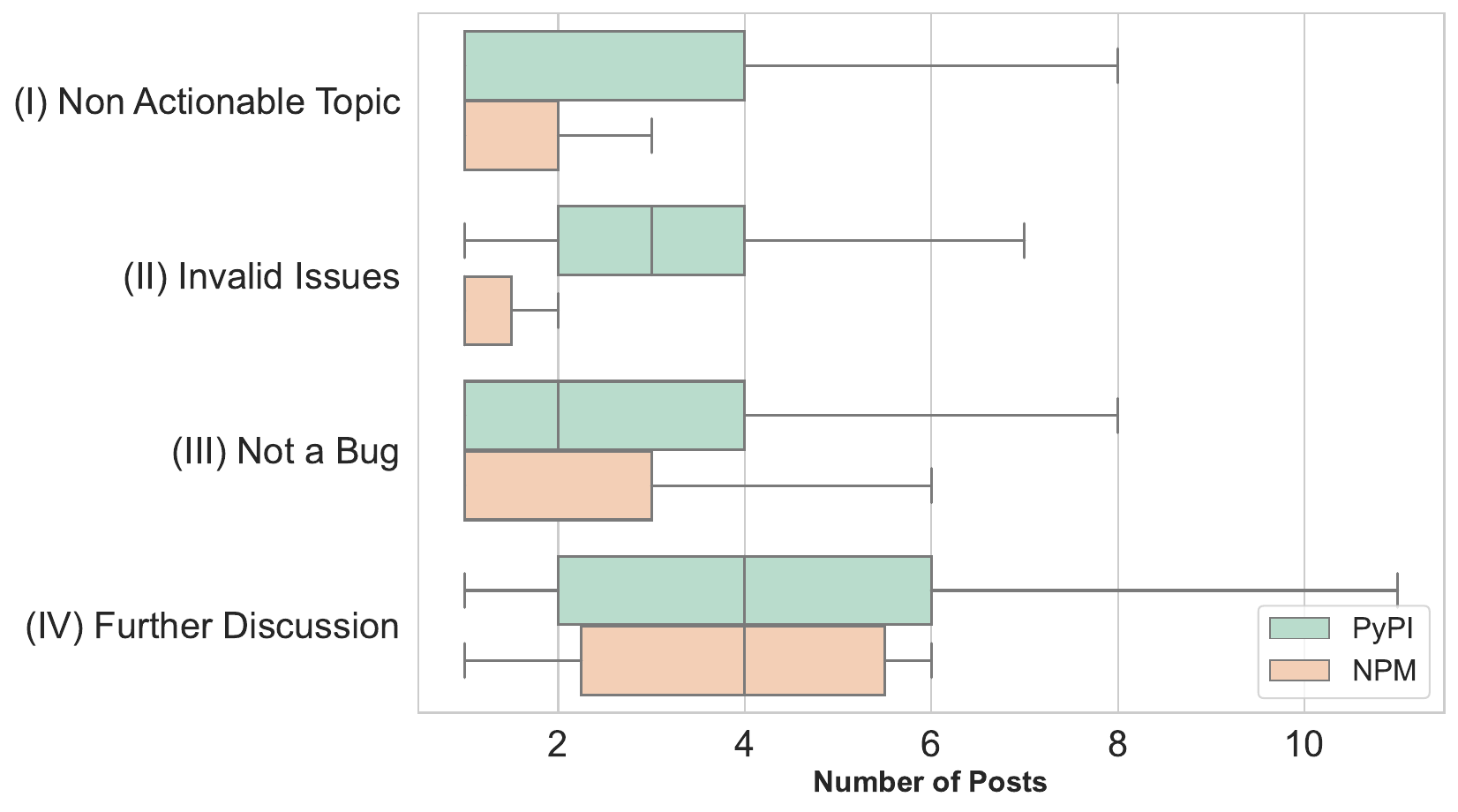}}
    \caption{Discussions that are converted from issues: (I) raising time and (II) number of posts until the raising time.}
    \label{fig:RQ3_b}
\end{figure}

To answer RQ3, we analyze the process of the conversion from discussion to issues and vice versa, in terms of the following two aspects: (I) the raising time of conversion intent and (II) the number of post until the raising time.
We below present the two metric related results (Figure~\ref{fig:RQ3_a} and Figure~\ref{fig:RQ3_b}) with regard to two conversion kinds.

\textbf{(I) Conversion from discussion to issue.}
Figure~\ref{fig:RQ3_a} presents the results of the computed metrics for the discussions that are suggested to be converted into issues.
As shown in Figure~\ref{fig:RQ3_a} (a), on the one hand, we find that two ecosystems share a similar raising time in terms of categories \textit{Reporting an Enhancement} and \textit{Reporting a Bug}.
For example, we observe that \textit{Reporting an Enhancement} category takes a relatively long time (i.e., the medians of raising time are 74 hours and 67.8 hours for the NPM and the PyPI, respectively) when the conversion intent is raised when compared to the other three reason categories.
Such a result indicates that it may take time for developers to discuss and reach a consensus to propose an enhancement in the issue tracker.
On the other hand, significantly large differences across two ecosystems are observed in terms of categories \textit{External Repository} and \textit{Reporting a Clarification Request}, validated by the Mann-Whitney test with \textit{p-value} $<$ 0.05.
For instance, \textit{External Repository} category takes the least time to be identified for the NPM, i.e., the median of raising time is 3.9 hours.
While, for the PyPI, it takes almost a median of 69.5 hours to be notified of the external repository.
One possible reason is that the NPM packages are less likely to be isolated when compared to the PyPI ones~\citep{decan2016topology}.
For the metric related to the number of posts, as shown in Figure~\ref{fig:RQ3_b} (b), the median posts of \textit{Reporting an Enhancement} category for both ecosystems are two which are larger than another three reason categories, suggesting that more discussions are likely to be involved in this instance.
However, the Mann-Whitney test suggests that there is no significant difference for all the paired categories between the NPM and the PyPI.

For the statistical test, Kruskal-Wallis H tests confirm that the hypothesis \textit{`Raising time and the number of posts until the conversion intent are significantly different among reasons of the conversion'} is established in terms of the conversion from discussions to issues, with \textit{p}-value~\textless~0.001 for the \textit{Raising time} and \textit{p}-value~\textless~0.05 for the \textit{Number of posts} for the NPM repositories. However, the hypothesis is not supported for both raising time and the number of posts within the PyPI repositories.

\textbf{(II) Conversion from issue to discussion.} 
Figure~\ref{fig:RQ3_b} shows the results of the computed metrics for the discussions that are suggested to be converted from issues.
Note that we only analyze the relatively frequent reasons for the two studied ecosystems (i.e., those instances whose frequencies are greater than 10 in Table~\ref{RQ1:result}).
As shown in Figure~\ref{fig:RQ3_b} (a), we observe that it takes relatively a long time to receive the convention intent, i.e., the median is around 15.2 hours and 35.1 hours for all reason codes within the NPM and the PyPI, separately.
More specifically, 17.5 hours and 24.5 hours (median) are taken for \textit{Non Actionable Topic} for the NPM and the PyPI, respectively.
At the same time, we observe that there exists a significant difference via the Mann-Whitney test in terms of the category \textit{Invalid Issues} between the two ecosystems.
It takes a much longer time for developers to raise this intent in the PyPI, i.e., a median of 79 hours.
On the other hand, as shown in Figure~\ref{fig:RQ3_b} (b), few posts are involved until the conversion intent of \textit{Non Actionable Topic} is raised, with the median being one for both ecosystems.
This result suggests that this conversion intent is likely to be raised in the first post of an issue.
Unsurprisingly, \textit{Further Discussion} category involves the most posts (i.e., the median is four for two ecosystems).
Furthermore, the Mann-Whitney test confirms that there is a significant difference in categories \textit{Non Actionable Topic} and \textit{Invalid Issues} between the NPM and the PyPI.
This suggests that relatively more posts are submitted to decide these two conversions with the PyPI. 

For the statistical test, Kruskal-Wallis H tests confirm that the hypothesis \textit{`Raising time and the number of posts until the conversion intent are significantly different among reasons of the conversion'} is established for the NPM repositories in terms of the conversion from issues to discussions, with \textit{p}-value~\textless~0.05 for both the \textit{Raising time} and the \textit{Number of posts}. 
For the PyPI repositories, a significant difference is observed for the \textit{Number of posts} but is not observed for the \textit{Raising time}.

\begin{tcolorbox}[colback=gray!5,colframe=gray!75!black,title= RQ3 Summary]
Our results show that there are few posts involved until the conversion intent is raised, i.e., mostly the median is one in both studied conversion kinds.
However, we observe that deciding on a conversion may potentially take time. For instance, the median time of the conversion from issues to discussions is 15.2 hours and 35.1 hours for the NPM and the PyPI ecosystem, respectively.
\end{tcolorbox}


\section{Implications}
\label{sec:discussion}

Based on the results of our RQs, we now discuss the implications of the study and provide suggestions accordingly.

\smallskip
\noindent
\textbf{Reduce unnecessary conversion in the first place.} 
Findings from the study illustrate a variety of reasons behind conversions, i.e., four reasons of converting discussion to issues (RQ1) and seven reasons of converting issues to discussions (RQ2).
GitHub guidelines pointed out the hint that the topics like questions and ideas should be raised in the GitHub Discussion.
While, we also find that other reasons exist, for instance, \textit{invalid issues} and \textit{not a bug}.
These findings could complement the GitHub Discussion guidelines to guide those repositories that intend to adopt the GitHub Discussion feature.
To maintain these channels, our constructed taxonomy can help maintainers and contributors decide when to contribute to each channel. 

We further performed an additional analysis to investigate whether or not the inconsistent use of these two channels exists.
In the context of feature requests (between \textit{Reporting an Enhancement} and \textit{Non Actionable Topic}),
first, we observe that inconsistency does exist in the specific repositories. 
For example, within a repository namely \textit{next.js}, on the one hand, a discussion\footnote{https://github.com/vercel/next.js/discussions/12325} that requested adding custom attributes was suggested to be opened as an issue.
On the other hand, an issue that requested a configuration setting was converted into a discussion and the author expressed confusion ``Hey team, why did a feature request become a discussion?...''. Similarly, another author from the discussion\footnote{https://github.com/vercel/next.js/discussions/27756} doubted ``Is it normal that feature requests are moved to a discussion?''.
Second, we notice that different repositories tend to have their own rules.
For example, a collaborator from the \textit{superset} repository left a comment\footnote{https://github.com/apache/superset/discussions/19185} ``Moving this feature request to Github Discussions so we can keep Issues focused on bugs!'', while another repository \textit{ant-design} allows feature requests to be proposed as one maintainer suggested in a discussion\footnote{https://github.com/ant-design/ant-design/discussions/29818} ``Since this not a bug or feature request. Move to discussion instead.''. 
Such inconsistent use is also observed in the cases where issues should be created until enough information is provided or not (between \textit{Reporting a Clarification Request} and \textit{Invalid Issues}).
Given the ecosystem PyPI, we observe that 28 out of 32 instances classified into \textit{Invalid Issues} are from the \textit{airflow} project. This may indicate that this project could use discussions as a gatekeeping mechanism. 
At the same time, we also notice that inconsistent use of two different channels did exist in the \textit{airflow} project. For example, in this instance\footnote{https://github.com/apache/airflow/discussions/14315}, the contributor suggested the author to open an issue from the discussion thread with some replication information (i.g., DAG).
Based on these insights, to relieve the confusion for contributors and improve the user experience, we suggest that project maintainers should clearly state the submission rules in their project README or contribution guidelines. 

A side-effect of keeping separate channels is duplication.
Our empirical observations are align with the survey insights where the developers face the challenge of the duplication~\citep{Hata2022GitHubDA}.
We find that duplication indeed exists either between GitHub Discussions, or between GitHub Discussions and Issues.
In terms of duplication between GitHub Discussions and Issues, for example, in this discussion~\footnote{https://github.com/invertase/react-native-firebase/discussions/4290}, the maintainer left the post and pointed out that \textit{``Already tracking here \#4283 - please check issues first''}.
Duplicate software artifacts have been proven to cause additionally unnecessary efforts and reduce efficiency.
Many techniques are proposed to detect duplicate artifacts using information retrieval, such as issue report~\citep{nguyen2012duplicate, hindle2016contextual}, pull request~\citep{li2017detecting, wang2019duplicate} on GitHub.
We notice that the latest work has started to explore the related posts (duplicate or near duplicate) in GitHub Discussions and proposes an approach based on a SentenceBERT pre-trained model~\citep{lima2022looking}.
Hence, another challenge would be the detection and removal of such duplication between these different channels.

Furthermore, the study shows evidence that GitHub Discussion not only facilitates communication but also can lead to actionable contributions to the project. 
For instance, our manual classification in RQ1 shows that these contributions are also significant, i.e., 76.6\% of samples being classified for \textit{Reporting a Clarification Request}, \textit{Reporting a Bug}, and \textit{Reporting an Enhancement} reason categories for the NPM ecosystem (Table \ref{fig:rq2_reason}).
We point out that maintainers  may consider Discussion as a means to attract and onboard potential new contributors, as Discussion can be acted as an incubation for those uncertain problems (e.g., potential bugs). After the evaluation by the project developers, these initial discussions have opportunities to be converted into actionable contributions.

\smallskip
\noindent
\textbf{Conversion is not trivial.} 
Through the manual analysis of RQ2, our results show that some conversions may not be trivial to be decided.
For instance, we observe that \textit{Not a Bug} reason category accounts for up to 20.5\% and 18.4\% for the two ecosystems (Table \ref{RQ1:result}). 
This indicates that the bug proposed by the contributor is not true from the engineering developer's perspective and thus it is converted into a discussion.
We argue that such a type of conversion is not trivial and would take up time and pipeline for the developers to investigate the real causes.
For instance, in the example\footnote{https://github.com/renovatebot/renovate/discussions/14457}, the contributor originally submitted an issue (including bug description and bug related logs).
While, after the confirmation, the maintainer commented that \textit{``This is not a bug. It's working as intended. Please open a discussion first. Maybe we can allow this as a feature request.''}.
Based on this insight, we suggest the contributor, especially those newcomers, to start from the GitHub Discussion for the uncertainty problem.
At the same time, one potential future research direction would include the tendency analysis of non-trivial conversions.
Some conversions would be converted back and forth, thus it is valuable to know what kinds of discussions or issues are difficult to get a consensus, especially bug related topics.
In addition, the fact exists that invalid issues could be closed while they are not converted to Discussion.
Hence, another direction for researchers is to investigate how common it is and the reasons behind this fact, which would gain a deeper insight into the maintenance of channels.

\smallskip
\noindent
\textbf{Raising conversion intent potentially takes time.}
In RQ3, a quantitative analysis was conducted to look at the conversion process.
Although few discussions are involved, the time to receive the conversion intent is relatively long (Figure~\ref{fig:RQ3_a} and Figure~\ref{fig:RQ3_b}).
On the one hand, for those non-trivial reasons (e.g., \textit{Reporting an Enhancement} and \textit{Not a Bug}), we argue that such a long process would not be avoidable since these discussion/issue topics may require time and conversations to confirm their significance and correctness, regardless of the channel it happens on.
Thus, we recommend developers that they should accommodate the development pipeline of the contributed projects.
On the other hand, for those trivial reasons (e.g., \textit{Non Actionable Topic} and \textit{External Repository}), we find that it still takes almost 17.5 hours and 24.5 hours for the issues related to non actionable topics to be converted into discussions for the NPM and the PyPI.
One potential reason could be that these topics are not being paid attention to by the community members.
Hence, a future direction for the researchers is to investigate whether the authorships of discussions or issues play a role in the conversion process.
Interestingly, our statistical tests indicate that raising conversion intent within the PyPI likely takes a longer time and involves more posts than the NPM in some specific categories, e.g., \textit{External Repository}, \textit{Invalid Issues}.
Therefore, we encourage researchers to further explore how the nature of the ecosystem will affect the conversion process.
Meanwhile, we argue that such a long process of trivial ones could be wasteful, hence another direction is to call for a promising classifier that is able to identify these non actionable topics effectively to make the issue tracker lighter and save the developers' efforts.

\section{Threads to Validity}
\label{sec:threat}
Below, we describe the threats to the validity of this study:

\paragraph{\textbf{External validity.}} External validity is with regard to the ability to generalize based on our results. In this study, we conduct a case study of NPM package repositories (web libraries) that adopt the Discussion feature. 
Thus, the observations based on this case study may not be generalized to other kinds of repositories (e.g., system software).
However, our goal is not to build a theory to fit all repositories, but rather to shed light on the challenge that developers face during the choice between GitHub Issue and Discussion.
Nonetheless, additional replication studies would help to generalize our observations in other repository domains, such as software tools, application software, and Non-web libraries and frameworks.
Thus, to encourage future replication studies, we disclose our research materials and provide a replication package online, including raw NPM repository discussion data, manually labeled data, and the script to retrieve the discussions. 
The package is available: \url{https://github.com/posl/GitHub_Discussion_Conversion}.

\paragraph{\textbf{Construct validity.}} Construct validity refers to the degree to which our measurements capture what we aim to study.
During the data collection of those discussions that should be converted into issues, we rely on the heuristic approach using a list of keywords (e.g., open, transfer, convert, and so on) to filter the discussion posts.
The threat may occur due to the incompleteness of the initialized keyword list.
To mitigate this threat, we manually explored the random 100 discussion posts, and inspected the potential keywords that were highly frequent.
At the same time, the goal is not to retrieve all these kinds of discussions, instead, we aim to get insights from a sufficient sample size. 
Thus, we believe that our sample size is sufficient enough to provide insightful observations.

\paragraph{\textbf{Internal validity.}} Internal validity denotes the approximate truth about inferences regarding cause-effect or causal relationships. Three threats are summarized.
First of all, to understand the reasons behind the conversion, we performed a manual analysis, which may be mislabeled due to its subjective nature.
To relieve such a threat, we conducted the manual coding in multiple iterations with two authors and calculated the Kappa agreement scores to ensure the quality.
Once the scores suggested nearly perfect or substantial agreement, the first author then coded the rest of the samples.
Additionally, to separate category \textit{Reporting a Bug} and category \textit{Reporting a Clarification Request}, we refer to whether the possibility-related word is given in a comment.
This is likely to introduce a bias due to the natural usage of English words.
The second threat occurs in the metric computation (RQ3). 
Since there is no automatic method to identify the exact time of the conversion, we rely on the post time when the conversion intent is raised.
The third threat may exist in the selection of statistical tests.
To test the significance of the studied metrics among different conversion reasons (RQ3), we use the  Kruskal-Wallis H test.
The cause-effect may differ from the other statistical tests.
We are however confident, as the selected test is widely used in the prior study~\citep{chinthanet2021lags, wang2021automatic}. 

\section{Related Work}
\label{sec:related}
In this section, we position our work with respect to the literature on question and answer forum, developer communication in software development, and barrier for newcomers in OSS projects.

\subsection{Question \& Answer Community}
Developers often turn to programming question and answer (Q\&A) communities to seek for help with their codes.
Stack Overflow, as one of the most popular Q\&A communities~\citep{wang2023understanding}, has become a gold mine for software engineering research and is found to be useful for software development. 
\cite{chris_2011}, as a pioneer work, categorized the kinds of questions that are asked, and explored which questions are answered well and which ones remain unanswered. 
\cite{vasilescu2012gender} provided a quantitative study of the phenomenon, in order to assess the representation and social impact of gender in Stack Overflow.
\citet{treude2017understanding} conducted a survey-based study to understand developers’ information needs as they relate to code fragments.
A large body of studies targets the knowledge extraction and challenges of specific topics from Stack Overflow.
To facilitate program repair, \cite{liu2018mining} proposed an approach to extract code samples from Stack Overflow, and mined repair patterns from extracted code samples. 
\citet{wan_tse_2021} used Stack Overflow
to understand the challenges and needs amongst blockchain developers by applying Balanced LDA.
\citet{bangash2019developers} studied the machine learning related posts and found that some
machine learning topics are significantly more discussed than others, and others need more attention.

To brainstorm feature ideas, help new users get their bearings, and further improve collaboration, GitHub released Discussion in 2020. 
\citet{Hata2022GitHubDA} took a first look at the early adoption of GitHub Discussion from several aspects. 
For instance, they found that errors, unexpected behavior, and code reviews are prevalent discussion categories.
Specifically, the perception from the developer survey pointed out that developers consider GitHub Discussions useful but face the problem of topic duplication between Discussions and Issues.
Motivated by their survey insight, we conduct an empirical study on NPM repositories to understand how developers maintain these channels, so as to complement the knowledge gap and provide empirical suggestions for the practitioners on how to select the appropriate channel.

\subsection{Developer Communication in Software Development}
Developer communication plays a significant role in software development, such as code review process~\citep{WANG2021111009}.
\citet{Bacchelli_ICSE2013} stated that developers have need of richer communication than comments annotating the changed code when reviewing. 
\citet{Information_2018} found that during reviews, reviewers often request additional information about
correct understanding, alternative solution, to improve patch quality.
Recent work cited that reviewers suffer from confusion due to a lack of information about the intention of a patch~\citep{confusion_saner_2019}. 
\citet{WANG_emse} observed that developers are likely to share links during review discussions with seven intentions to fulfill information needs.
Meanwhile, \citet{hirao2019fse} reported that the patch linkage (i.e., posting a patch link to another patch) is used to indicate patch dependency, competing solutions, or provide broader context.
In addition to code reviews, nowadays interactive communication channels are available to support development.
\citet{slack_jss} conducted a longitudinal study on coordination to understand the use of meetings and Slack and found that collaboration tools increase awareness and informal communication.
\citet{parra2022comparative} presented a comparative study of developer communications on Slack and Gitter.
Raglianti et al. (2022) proposed a tool using Discord conversations to aid program comprehension.
With the official release of GitHub Discussion, it would gradually become a popular centre for developers to communicate and share ideas during their software development. 
Hence, one potential benefit of our study is to improve the communication efficiency between GitHub Discussion and Issue.

\subsection{Barriers for Newcomers in OSS Projects}
Newcomers are significant to the survival, long-term success, and continuity of OSS projects~\citep{Kula2019book}.
However, literature pointed out that newcomers face many challenges in their initiative activities in OSS projects~\citep{Steinmacher2014}. 
For instance, \citet{lee2017understanding} reported that newcomers lack the necessary domain knowledge and programming skills.
In addition, some non-technical also affect newcomers' onboarding process, such as communication and social interaction~\citep{tan2019communicate, rehman2022newcomer}. 
In the recent work, \citet{mendez2018open} studied newcomer barriers and gender through a new perspective, i.e., the usage of OSS tools and infrastructure.
To facilitate the newcomers' onboarding process, a series of theories and strategies have been proposed.
\citet{steinmacher2018let} provided guidelines for
both OSS communities interested in receiving more external contributions, and newcomers who want
to contribute to OSS projects.
\citet{first_issue_2020} discovered the criteria to identify good first issues (GFIs) that may make GFIs
more likely to be solved by newcomers.
In the following study, \citet{xiao2022recommending} proposed RECGFI, an effective practical approach for the recommendation of good first issues to newcomers.
With the introduction of a new feature (i.e., GitHub Discussion), the newcomers may also face the barrier of choosing proper communication channels when asking questions or raising issues.
Especially for those uncertain problems (i.e., whether they are real issues or not), our results suggest that it would be appropriate for newcomers to start from the Discussion channel.

\section{Conclusion}
\label{sec:conclusion}
With the adoption of the GitHub Discussion feature, it becomes challenging for developers to appropriately choose or maintain between GitHub Discussion and Issue. 
In this work, we conducted an empirical study on 259 NPM and 148 PyPI repositories to understand the reasons behind converting Discussion to Issue and vice versa, and to investigate whether or not the conversion requires additional effort. 
Our empirical results show that reporting a clarification request is the most common reason of converting discussions to issues (35.1\% and 34.7\%, respectively), while having non actionable topic is the most frequent reason of converting issues to discussions (55.0\% and 42.0\%, respectively).
Moreover, we observe that it potentially takes time to raise a conversion intent. 
This study contributes to helping developers effectively
utilize these different communication channels, and also provides the future direction on the automatic classifier needs such as duplication detection between GitHub Discussion and Issue, and identification of non actionable topics from Issue.

\section*{Acknowledgement}
This work is supported by Japanese Society for the Promotion of Science (JSPS) KAKENHI grants (JP20K19774, JP20H05706, JP22K17874, JP21H04877, JP23K16864), and JSPS and SNSF for the project ``SENSOR'' (JPJSJRP20191502).

\section*{Declarations}

\subsection*{Conflict of Interest}
The authors declare that Raula Gaikovina Kula and Yasutaka Kamei are members of the EMSE Editorial Board. 
All co-authors have seen and agree with the contents of the manuscript and there is no financial interest to report.

\subsection*{Data Availability Statements}

The datasets generated during and/or analysed during the current study are available in the \url{https://github.com/posl/GitHub_Discussion_Conversion}.

\bibliographystyle{spbasic}
\bibliography{myref}
\end{document}